# A Lateral AlGaN/GaN Schottky Barrier Diode with 0.36 V Turn-on Voltage and 10 kV Breakdown Voltage by Using Double Barrier Anode Structure

Ru Xu, Peng Chen, Jing Zhou, Yimeng Li, Yuyin Li, Tinggang Zhu, Dunjun Chen, Zili Xie, Jiandong Ye, Xiangqian Xiu, Rong Zhang, *Member, IEEE*, Youdou Zheng

*Abstract*—In this letter, we demonstrate a lateral AlGaN/GaN Schottky barrier diode (SBD) on sapphire substrate with low turn-on voltage ($V_{on}$) and high breakdown voltage ($V_{BK}$). By using a double barrier anode (DBA) structure formed by the mixture of Platinum (Pt) and Tantalum (Ta), the $V_{on}$ of the SBD can be as low as 0.36 V with a leakage current of $2.5\times10^{-6}$ A/mm. Supported by the high-quality carbon-doped GaN buffer on sapphire, the $V_{BK}$ can reach more than 10 kV with the anode-to-cathode spacing of 85 μm. Combining the $V_{BK}$ and the specific on-resistance ($R_{on,sp}$) of 25.1 mΩ·cm$^2$, the power figure of merit ($V_{BK}^2/R_{on,sp}$) of the SBD can reach 4.0 GW/cm$^2$, demonstrating a great potential for the application in ultra-high-voltage electronics.

*Index Terms*—lateral AlGaN/GaN Schottky barrier diode, double barrier anode, turn-on voltage, breakdown voltage.

## I. Introduction

AlGaN/GaN structure can generate two-dimensional electron gas (2-DEG) with high electron concentration (~$1\times10^{13}$ cm$^{-2}$) and high electron mobility (~2000 cm$^2$/V·s). Combined with the high critical electric field of GaN (~3.3 MV/cm), AlGaN/GaN-based power electronic devices can have faster switching speeds, lower on-resistance and higher breakdown voltages than Si-based devices. In all of them, the power Schottky barrier diode (SBD) is one of the important devices, which can meet the requirements of various scenarios such as high voltage, high frequency switching, high temperature and high power density. They have a wide range of application prospects in consumer electronics, automotive electronics, new energy, industrial motors and even to ultra-high-voltage (UHV, >10 kV) electronics.

In recent years, a large number of high-performance AlGaN/GaN SBDs have been reported, with turn on voltages ($V_{on}$) of 0.2 V~0.8 V and breakdown voltages ($V_{BK}$) of 0.13 kV~10.0 kV. Apparently, low $V_{on}$, low specific on-resistance ($R_{on,sp}$) and high $V_{BK}$ with low leakage current ($I_{leakage}$) are essential for high-performance AlGaN/GaN power SBDs. However, $V_{on}$, $I_{leakage}$ and $V_{BK}$ are mainly determined by the Schottky contact, which is not easy to improve these three parameters at the same time. In previous reports, in order to achieve low $V_{on}$, anode recess structure has been widely used [1-8]. The $V_{on}$ can be further reduced to < 0.5 V with a low work function anode metal, such as W [6, 8], or with ohm-Schottky combined anode [1-4], which may also bring some negative impact on $I_{leakage}$ and $V_{BK}$. However, simply using a low work function metals as the Schottky electrode brings some disadvantages, such as higher leakage current and lower reverse breakdown voltage. Therefore, how to use low work function metals while maintaining the high breakdown voltage of the Schottky electrodes requires a new electrode structure design. For the optimization of $V_{BK}$, high-voltage SBDs with various terminal structures have been demonstrated [9,17,22-27], including the 2.7-kV and 3.4-kV AlGaN/GaN SBDs with field plates (FP) reported in our previous work [28,29]. Although the FP structure is adopted, we found that it is still difficult to significantly increase the $V_{BK}$ of AlGaN/GaN SBD on Si substrate for ultra-high voltage (UHV, >10kV) [30]. By using high-quality GaN materials on SiC or sapphire substrates, high-performance AlGaN/GaN SBDs with $V_{BK}$ of > 9 kV and >10 kV were reported [14,22,25,26,30]. Obviously, a proper anode structure for the $V_{on}$ < 0.5 V and a high-quality GaN material for the $V_{BK}$ >10 kV are the key factors to realize high-performance AlGaN/GaN SBDs, which are still being explored now.

In this work, a double barrier anode (DBA) structure composed of high work function (Pt, 5.65 eV) and low work function (Ta, 4.25 eV) was proposed, supported by a high-quality carbon-doped GaN buffer layer on sapphire

This work is supported by National High-Tech Research and Development Project (2015AA033305), Jiangsu Provincial Key Research and Development Program (BK2015111), Collaborative Innovation Center of Solid State Lighting and Energy-saving Electronics, the Research and Development Funds from State Grid Shandong Electric Power Company and Electric Power Research Institute. (*Corresponding authors: Peng Chen; Rong Zhang.*)

R. Xu is with the Key Laboratory of Advanced Photonic and Electronic Materials, Nanjing University, Nanjing 210093, China and also with the Nanjing University of Information Science and Technology, Nanjing 210044, China.

P. Chen, J. Zhou, Y.M. Li, Y. Y. Li, D.J. Chen, Z. L. Xie, J. D. Y, X. Q. Xiu, R. Zhang, Y.D. Zheng are with the the Key Laboratory of Advanced Photonic and Electronic Materials, Nanjing University, Nanjing 210093, China (Corresponding e-mail: pchen@nju.edu.cn; rzhang@nju.edu.cn).

T.G.Zhu is with CorEnergy Semiconductor Co. Ltd, Suzhou 215000, Chnia.

substrate, we successfully fabricated the AlGaN/GaN SBD with low $V_{on}$, low $I_{leakage}$ and ultra-high $V_{BK}$.

## II. Device Fabrication

The samples were grown by metal organic chemical vapor deposition (MOCVD) on 2-inch c-plane sapphire substrate. From the substrate, the device structure consists of a nucleation layer, a 3-μm carbon doped GaN buffer layer, a 200-nm i-GaN channel layer, a 1-nm AlN spacer, a 20-nm Al$_{0.25}$Ga$_{0.75}$N barrier layer, a 2-nm GaN cap layer and a 50-nm in-situ SiN$_x$ layer.

First, the photolithography defines the SBD isolation area, then the SiN$_x$ was etched out by reactive ion etching (RIE) using the gas of CF$_4$, and the isolation mesa was etched out by inductively coupled plasma (ICP) using the mixture of Cl$_2$/BCl$_3$. In the second step, the anode recess area was also etched by ICP using the mixture of Cl$_2$ and BCl$_3$. Then the sample was etched in a diluted KOH solution for 15 minutes at 80 °C to remove the etching damages. The last step, we the mixture of Ta (50 nm) and Pt/Au (50/300 nm) as anode which presents a serrated shape as shown in Fig.1. The angle between each piece of Ta is 22.5 degree, and the Ta edge is designed to be round to avoid premature damage. The cathode is Ti/Al/Ni/Au (30/150/30/100 nm) by rapidly annealing at 850 °C for 30 s in N$_2$. SBDs with single Pt/Au (50/300 nm) and Ta/Au (50/300 nm) anode were also fabricated. All SBD devices in this work feature a circular anode with a radius of 90 μm and an anode-cathode spacing ($L_{AC}$) of 85 μm. The overhang length of Ta beyond the anode recess is 2 μm and the overhang length of Pt/Au beyond the anode recess is 3 μm. There are no field-plates or any other terminal structures in our devices. The extracted ohmic contact resistance ($R_C$) and sheet resistance ($R_{SH}$) were 0.62 Ω·mm and 310 Ω/□, respectively.

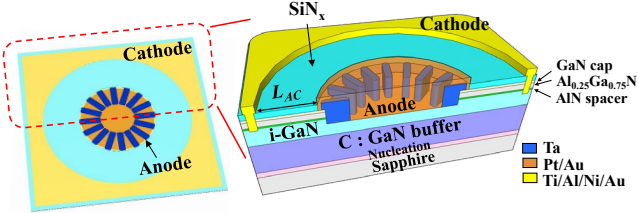

Fig.1. Top and Cross-section view of the double barrier anode AlGaN/GaN SBD.

The forward I-V characteristics were measured using a Keithley 4200 semiconductor characterization system, the reverse I-V characteristics were measured using an IWATSUCS12105C semiconductor curve tracer.

## III. Results And Discussions

Fig 2. (a) and (b) show the forward and reverse I-V characteristics of SBDs with different anode structures, including the DBA structure and the traditional anode structures with single anode metal Pt or Ta. The specific performances extracted from Fig. 2. (a) and (b) are shown in

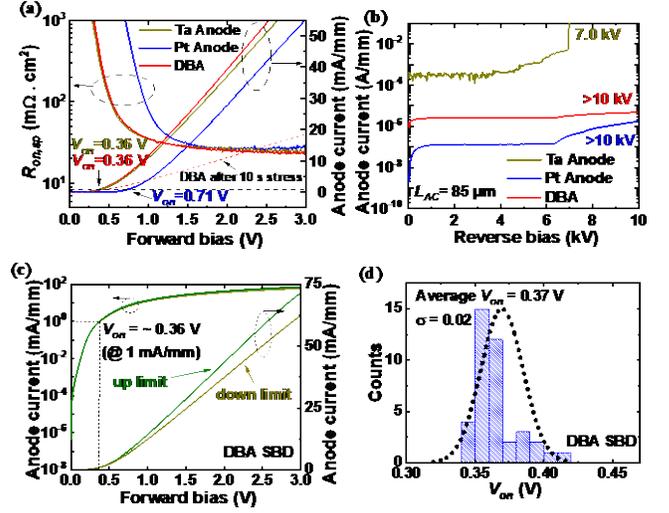

Fig.2. (a) Forward (b) Reverse I-V characteristics of SBDs with different anode structures. (c) Forward I-V characteristics and (d) turn-on voltage distribution of 40 DBA SBD devices ($L_{AC}$=85 μm).

TABLE I
COMPARISON OF THE KEY PERFORMANCES OF SBDS WITH DIFFERENT ANODE STRUCTURES.

| Anode | $V_{on}$ (V) | $R_{on,sp}$ (mΩ·cm$^2$) | $I_{leakage}$ (A/mm) | $V_{BK}$ (kV) |
|---|---|---|---|---|
| Ta | 0.36 | 25.5 | 9.2×10$^{-4}$ | 7.0 |
| Pt | 0.71 | 25.8 | 1.3×10$^{-7}$ | >10 |
| DBA | 0.36 | 25.1 | 2.5×10$^{-6}$ | >10 |

Table I. This paper emphasizes the intrinsic properties of GaN material, so the breakdown voltage is taken from the physical breakdown. The $V_{on}$ was defined as the forward voltage at 1 mA/mm, the $R_{on,sp}$ was calculated under the consideration of a 1.5-μm transfer length of ohmic contact and a 1.5-μm extension length of the Schottky contact, and the $I_{leakage}$ was extracted at -5.0 kV. As shown in Table I, for the forward characteristics, the $V_{on}$ (at 1 mA/mm) values of the DBA device and Ta anode device are both 0.36 V, nearly half of the Pt anode device. As shown in Fig.2(c) and (d), we counted the results of 40 DBA SBDs. The $V_{on}$ distribution range is 0.35 V ~ 0.42 V, the everage value is 0.37 V and the dominant value is 0.36 V. By estimating a total transfer length $L_T$ of 3 μm for the calculation of the active area, the working voltage of the DBA device extracted at a forward current of 100 A/cm$^2$ is 3.7 V. The $R_{on}$s of ~29 Ω·mm for the three devices were obtained in Fig. 2. (a). By calculating $R_{on,sp}=R_{on}×(L_{AC}+L_T)$, the $R_{on,sp}$ is only a little different among the three devices. For the reverse characteristics, the $I_{leakage}$ of the DBA device is 2.5×10$^{-6}$ A/mm, which is one order of magnitude higher than that of the Pt anode device, but is two orders of magnitude lower than that of the Ta anode device. The $V_{BK}$ can reach more than 10 kV for DBA and Pt anode devices, but only 7.0 kV for Ta anode device. We tested more than five DBA SBDs, the $V_{BK}$ values are all higher than 10 kV. Furthermore, $V_{on}$ and $R_{on}$ degradation were measured after high voltage testing for the DBA SBDs. When the SBD was immediately turned on after high reverse voltage (8 kV) stress for 10 s, both $V_{on}$ and $R_{on}$ show some degradation. $V_{on}$ increases from 0.36 V to 0.49 V, and $R_{on}$ increases from 25.1 mΩ·cm² to 95.1 mΩ·cm², as shown in Fig.2 (a) as the red dotted line. However, this degradation is not permanent and can be fully recovered under a normal continuous forward voltage within a short period of time. The further optimization of the device structure is required, such as adding multi-field plate structure.

Based on the above experimental results, the power figure of merit (P-FOM=$V_{BK}^2/R_{on,sp}$) of DBA device was calculated to be 4.0 GW/cm². These performances indicate that the DBA structure can effectively reduce the $V_{on}$ of the device while maintaining low $I_{leakage}$ and high $V_{BK}$.

To reveal the specific mechanism of how the DBA structure works, a series of simulation studies were carried out by using Silvaco software. In the simulation, the anode was set as a DBA structure combined with the metal Pt and Ta, same to the actual device, Ta was also embedded in the peripheral Pt with a serrated shape. Fig.3 shows the electric field distribution of the SBD under different voltage conditions.

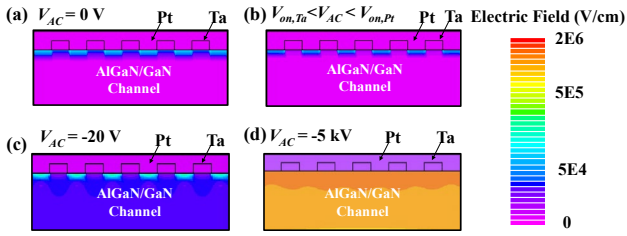

Fig.3 The electric field distribution of a DBA structure device under different voltage conditions. (a) $V_{AC}$=0 V (b) $V_{on,Ta}$< $V_{AC}$< $V_{on,Pt}$ (c) $V_{AC}$=-20 V.(d) $V_{AC}$=-5 kV.

For n-type Schottky contact, under ideal conditions, the depletion region width $W_d$ can be expressed as [31]

$$W_d = \sqrt{\frac{2\varepsilon_S(\varphi_B - V_n - V)}{qN_D}} = \sqrt{\frac{2\varepsilon_S(\varphi_m - \chi - V_n - V)}{qN_D}} \quad (1)$$

Where $\varepsilon_s$ is the relative dielectric constant of the semiconductor material, $\varphi_B$ is the Schottky barrier height, $V_n$ is the built-in potential at the bottom of the conduction band relative to the Fermi level, $V$ is the applied voltage, $q$ is the electron charge, $N_D$ is the donor concentration, $\varphi_m$ is the anode metal work function and $\chi$ is the semiconductor electron affinity. From this formula, it can be seen that $W_d$ increases with the increase of $\varphi_m$ of anode metal under ideal conditions. Then, the distribution of the electric field can reflect the formation of the depletion region [31].

When the anode voltage ($V_{AC}$) is 0 V, as shown in Fig 3. (a), the depletion region formed by Pt and Ta covered all area under the anode. Here, we use $V_{on,Ta}$ represent the $V_{on}$ of Ta anode, and $V_{on,Pt}$ represent the $V_{on}$ of Pt anode. When $V_{on,Ta}$<$V_{AC}$<$V_{on,Pt}$, as shown in Fig. 3. (b), the depletion region under Ta almost disappeared. At this stage, although the depletion region under Pt still exists, electrons can pass through the channel under Ta to turn on the SBD, which means the $V_{on}$ of the SBD is mainly determined by Ta. When $V_{AC}$<0 V, as shown in Fig. 3. (c), the depletion region formed by the peripheral Pt completely connected and become a continuous depletion region at a small reverse voltage of -20 V. At this time, a small number of electrons can pass through and generate leakage current compared to the single Pt anode device. When a large reverse bias of -5 kV is applied, as shown in Fig. 3(d), the entire depletion region is uniformly fully connected, and Pt almost dominates the depletion of the channel. The above results show that, according to the depletion region expansion behavior of GaN material, it can be realized under high reverse voltage that the role of Ta is mostly covered by Pt with the DBA structure to avoid the lower breakdown voltage from Ta. At the same time, since the Ta electrode only partially covers the anode, the leakage current due to the Ta electrode is also reduced.The leakage current of DBA device is one order of magnitude larger than that of the single Pt anode device, but two orders of magnitude lower than that of the Ta anode device, which also shows that Pt plays a dominant role under the reverse bias in the DBA device.

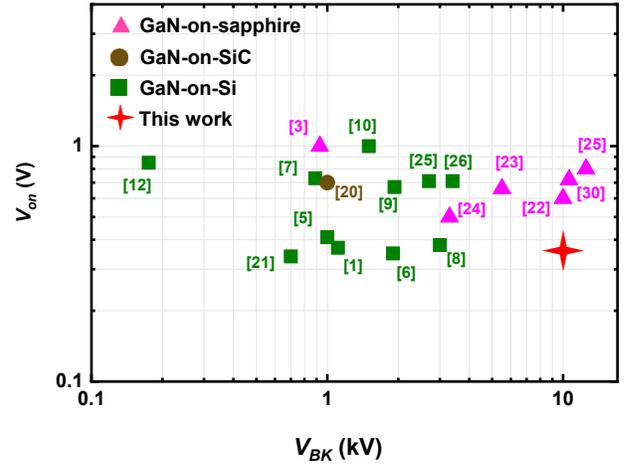

Fig.4 The benchmark of $V_{BK}$ veruse $V_{on}$ of GaN lateral SBDs on sapphire/SiC/Si substrate.

By combining the advantage of Ta under the forward bias and the advantage of Pt under the reverse bias, we use the DBA structure to realize the low $V_{on}$ and high $V_{BK}$ SBD. Fig. 4 plots the benchmark of $V_{BK}$ versus $V_{on}$ for GaN-based lateral SBD. The DBA SBD of this work not only showed the high $V_{BK}$ of > 10 kV among GaN-based lateral SBDs on any substrates, but also achieved the low $V_{on}$ of 0.36 V.

IV. CONCLUSION

In summary, in this work, by using the DBA structure and high quality GaN material on sapphire substrate, we realize the low $V_{on}$ and high $V_{BK}$ SBD. The $V_{on}$ of the DBA SBD is 0.36 V, while the $V_{BK}$ remains at 10 kV. Combined with the $R_{on, sp}$ of 25.1 mΩ·cm², the highest P-FOM of the device can be as high as 4.0 GW/cm². This work provides a feasible solution to achieve low turn-on AlGaN/GaN SBD together with ultra-high breakdown voltage, and is expected to realize the low-cost and high-performance applications of GaN materials in the field of UHV electronics.